\documentclass[aps,prb,twocolumn,superscriptaddress]{revtex4-1}
\usepackage{graphicx}
\usepackage[colorlinks=true, urlcolor=blue, pdfborder={0 0 0}]{hyperref}

\usepackage{amsmath}
\usepackage{subfigure}
\usepackage{epsfig} 
\usepackage{epstopdf}
\usepackage{float} 
\usepackage{tikz}
\usepackage[percent]{overpic}
\usepackage{comment}
\usepackage{placeins}
\usepackage{multirow}
\usepackage{rotating}
\usepackage{hhline}
\usepackage{xcolor}
\begin{document}
%\title{Cooling a many-spin system using a feedback control}
%\title{Adiabatic steering of a quantum system using a negative feedback control loop}

%\title{Polarizing and cooling many-spin systems using feedback control}

\title{Cooling classical many-spin systems using feedback control}

%\title{Cooling many-body systems using feedback control}

%\title{Control of many-body systems using negative feedback}

%\title{Quantum control of small spin clusters}

\affiliation{Department of Physics, School of Science and Engineering, The American University in Cairo, AUC Avenue, P.O. Box 74, New Cairo, 11835, Egypt}
\affiliation{Laboratory for the Physics of Complex Quantum Systems, Moscow Institute of Physics and Technology, Institutsky per. 9, Dolgoprudny, Moscow  region,  141700, Russia}
\affiliation{Institute for Theoretical Physics, University of Heidelberg, Philosophenweg 12, 69120 Heidelberg, Germany}

\author{Tarek A. Elsayed}
\email{tarek.elsayed@aucegypt.edu}
\affiliation{Department of Physics, School of Science and Engineering, The American University in Cairo, AUC Avenue, P.O. Box 74, New Cairo, 11835, Egypt}
\affiliation{Institute for Theoretical Physics, University of Heidelberg, Philosophenweg 12, 69120 Heidelberg, Germany}

\author{Boris V. Fine}
\email{fine.bv@mipt.ru}
\affiliation{Laboratory for the Physics of Complex Quantum Systems, Moscow Institute of Physics and Technology, Institutsky per. 9, Dolgoprudny, Moscow  region,  141700, Russia}
\affiliation{Institute for Theoretical Physics, University of Heidelberg, Philosophenweg 12, 69120 Heidelberg, Germany}
\affiliation{Institute for Theoretical Physics, University of Leipzig, Brüderstr. 16, 04103 Leipzig, Germany}

%\pacs{76.60.Pc, 02.30.Yy}

\date{\today}

\begin{abstract}

We propose a technique for polarizing and cooling finite many-body classical systems using  feedback control. The technique requires the system to have one collective degree of freedom conserved by the internal dynamics. The fluctuations of other degrees of freedom are then converted into the growth of the conserved one. The proposal is validated using numerical simulations of classical spin systems 
in a setting representative of Nuclear Magnetic Resonance experiments.
In particular, we were able to achieve 90 percent polarization for a lattice of 1000 classical spins starting from an unpolarized infinite temperature state.

\end{abstract}

%\pacs{76.60.Pc, 76.60.-k, 31.15.xv, 03.65.Sq, 75.10.Pq}
\maketitle

\section{Introduction}

Controlling the behavior of physical systems using feedback loops has been the subject of ever increasing interest both on the theoretical and experimental levels in the context of nuclear magnetic resonance (NMR) \cite{vandersypen2005,rugar05,degen2007,poggio13,li2017}, lasers \cite{machida, milburn93},  nanomechanical resonators \cite{schwab03,kleckner06,pinard,rossi2018}, trapped atoms and ions \cite{raithel02,steck04,zoller05} and other applications of quantum technology \cite{alexander05, bluhm, bartlett07, mahler, white10, wiseman2010,ashhab,clausen2010,clausen2012,hirose2016}, where concepts borrowed from the field of classical control have been applied\cite{sze}.

The present work exploits the potential of the feedback control to cool a thermally isolated {\it many-particle} system. This amounts to implementing a practical Maxwell demon to overcome the fundamental trend of entropy growth imposed by the second law of thermodynamics.  
We propose a feedback scheme, according to which a system is periodically driven  with an amplitude determined by a feedback loop from the measurement of a quantity that we want to steer to the desired value. Although periodic driving, normally, heats a many-body system \cite{Ji-18}, the proposed scheme does the opposite. While the idea behind the scheme is rather general, we focus specifically on spin systems in the context of NMR, where the experimental temperatures are, typically, very high on the energy scale of nuclear spins, which leads to very small nuclear polarizations and thus limits the use of NMR. 
Below we first present numerical simulations demonstrating that the scheme can, indeed, cool large finite spin lattices and then describe the mechanism, the limitations and possible generalizations of the scheme.

While our longer-term agenda is to develop a feedback scheme applicable to quantum systems, 
the scheme proposed in this work is validated only by direct classical simulations.  Direct quantum simulations in the presence of measurements are rather expensive computationally, which limits the size of the numerically accessible spin clusters.
The distinction between the quantum and the classical settings is that the dynamics is governed respectively by quantum spin commutators and classical spin Poisson brackets. According to detailed investigations of Refs.\cite{Elsayed-14,Elsayed-15,Starkov-18,Schubert2021}, the statistical averaging of classically simulated dynamics normally gives quantitatively accurate results for quantum observables such as the total magnetization.  The classical and quantum spin dynamics are known to exhibit qualitative differences as far as chaotic instabilities are concerned\cite{Elsayed-13-thesis,Fine-14,Elsayed-15q}, but these instabilities are not of principal importance in the present context.  

% {\it General formulation}
 \section{General formulation}
 
 We consider classical lattices of $N$ spins governed by the Hamiltonian ${\cal H} = {\cal H}_0 + {\cal H}_{\text{f}}$, where
 \begin{equation}
{\cal H}_0 =   \sum_{m<n}  \left[ 
J_{mn}^z S_m^z S_n^z + J_{mn}^{\perp} \left( S_m^x S_n^x+S_m^y S_n^y \right) \right]
\label{Ham-0}
\end{equation}
is the internal dynamics part, with $S_m^{\alpha}$ being the $\alpha$th projection of the $m$th classical spin having length $|\mathbf{S}_m| = 1$, $J_{mn}^{\alpha}$ are the interaction constants, $h_z$ is an external field and
\begin{equation}
{\cal H}_{\text{f}} = g(t) \sum_m S_m^x,
\label{Ham-f}
\end{equation}
is the feedback control term with
\begin{equation}
g(t)=g_0\cos(\omega t)\left[f(t)- M_z \right].
\label{g}
\end{equation}
Here $\omega$ is the driving frequency, $g_0$ is the amplitude prefactor, $M_z = \sum_m S_m^z$ is the total $z$-polarization of the system monitored by the feedback loop, and $f(t)$ is the function that steers $M_z$ with the goal of maximizing it. In the simulations below,  the steering function is simply linear: $f(t) = \dot{f} t$, where $\dot{f}$ is a time-independent parameter. When $f(t)$ entrains $M_z$, the latter also grows linearly on average, with small fluctuations near $f(t)$. Important for our scheme is the fact that, without driving, the Hamiltonian ${\cal H}_0$ conserves $M_z$, which means that $M_z$ stops changing whenever $M_z = f(t)$, and, in general, the closer $M_z$ to $f(t)$, the weaker the driving. Where relevant, the gyromagnetic ratios and the Boltzmann constant $k_B$ are set to 1.

The system is  initially at the infinite temperature state. At $t > 0$, it is thermally isolated from the environment and experiences only the dynamics governed by the Hamiltonian ${\cal H}$. The target of the feedback control is to produce a state with as low entropy as possible. The reduction of entropy can be due to the polarization of the system, which can be accompanied by the reduction of the energy $E \equiv \langle {\cal H}_0 \rangle$ (i.e. cooling).
The above formulation does not include any dissipation or decoherence due to an external environment, which means that the cooling should be achieved before the heating from the environment becomes prominent. In terms of solid-state NMR, this means that the time interval available for the feedback control is shorter than the spin-lattice relaxation time $T_1$.  

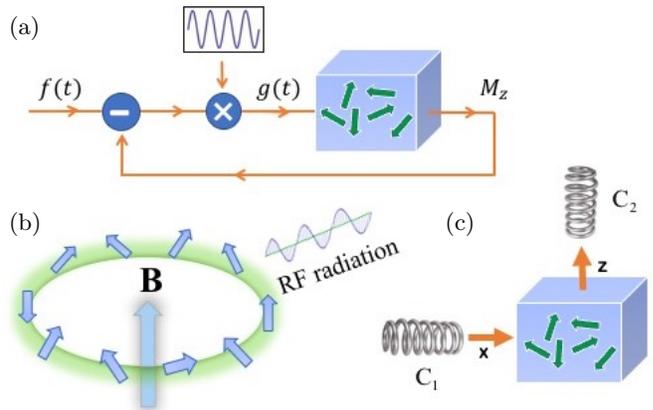
\begin{figure}[t!] \setlength{\unitlength}{0.1cm}
\begin{picture}(60 , 53 )
{
\put(-15, 0){\includegraphics[scale=0.5]{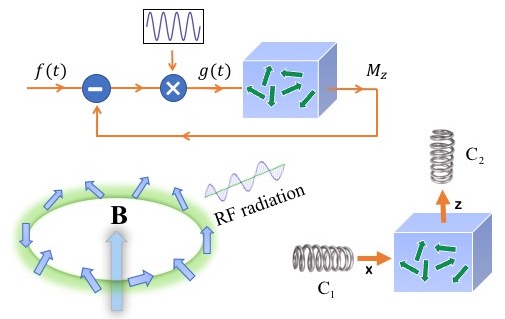}}

\put(-13, 51){\text{(a)}}
\put(-13, 25){\text{(b)}}
\put(45, 25){\text{(c)}}
}
\end{picture}
\caption{Illustrations of the proposed feedback scheme. (a) Block diagram: the system is driven by the periodic field $g(t)$ with amplitude modulated by the difference between the steering function $f(t)$ and the monitored quantity $M_z$. (b) NMR context: spins in a static magnetic field $\mathbf{B}$ driven by a modulated rf field. (c) Cartoon showing two principal elements required to implement the proposed scheme: source $C_1$ generating the rf field, and detector $C_2$ measuring $M_z$.}
\label{fig-concept}
\end{figure}

The block diagram of the proposed feedback scheme is shown in Fig.~\ref{fig-concept}(a).
In the context of NMR, with a strong static magnetic field oriented along the $z$-axis, the first sum in Hamiltonian ${\cal H}_0$ would represent the secular terms of nuclear spins-spin interaction in the Larmor rotated reference frame. The feedback control can then be implemented with the modulated radio-frequency (rf) magnetic field $\mathbf{g} = [g(t), 0, 0]$ acting on the spins along the $x$-axis [Fig.~\ref{fig-concept}(b)], while the second sum in ${\cal H}_0$ would represent the effect of a possible frequency offset from the resonance value. The implementation of the scheme conceptually requires two elements shown in Fig.~\ref{fig-concept}(c): one generating the feedback rf field (source $C_1$) and one measuring the monitored quantity $M_z$ (detector $C_2$). The actual implementation may not require coils but rather use, e. g., magnetic resonance force microscope \cite{rugar05,Issac-16,kovsata2020}, nitrogen-vacancy centers diamond \cite{Shagieva-19,bucher2019,cohen2020,dasari2021} or superconducting quantum interference devices (SQUIDs) \cite{McDermott-02}. Feedback from monitored fluctuations of finite nuclear spin clusters was used in \cite{rugar05,poggio13}, where the authors  created average polarization per spin of order $1/\sqrt{N}$.  Here, we aim at achieving the polarization per spin of order 1.

We note that even though $M_z$ is not the energy of the system, its increase implies lowering the entropy, which, in turn, can be easily converted into lower temperature, once, e.g., the system is placed in an external magnetic field.

%{\it Classical spins}
\section{ Numerical Simulation}
We now demonstrate that the scheme works for  $10\times 10\times 10 $ cubic lattice of classical spins with periodic boundary conditions and the interaction constants $J_{mn}^z = - 2 J_{mn}^{\perp} = \frac{\left(1-3\cos^2 \theta_{mn}\right)}{2 |\mathbf{r}_m-\mathbf{r}_n|^3}$, 
where $\mathbf{r}_m$ is the position of the $m$th lattice site and $\theta_{mn}$ is the angle between the $z$-axis and the vector $\mathbf{r}_m-\mathbf{r}_n$. The distance between the nearest lattice sites is equal to 1. The above choice of $J_{mn}^z$ corresponds to the truncated magnetic dipole interaction between nuclear spins in solids \cite{abragam,note-energy}. The parameters of the feedback control were:
$\omega = 7$, $g_0=0.2$ and $\dot{f}(t)=-0.005$. The simulations were based on solving the equations of motion for spin vectors 
$  \dot{\bf{S}}_m=  \bf{S}_m \times \bf{h}_m(t)$, where $\bf{h}_m(t)$ represents the local field at spin $m$, due to the interaction with the rest of the lattice and with the feedback field\cite{Elsayed-15}.  

\begin{figure}[t!] \setlength{\unitlength}{0.1cm}
\begin{picture}(60 , 46 )
{

\put(-16, 0){ \epsfig{file=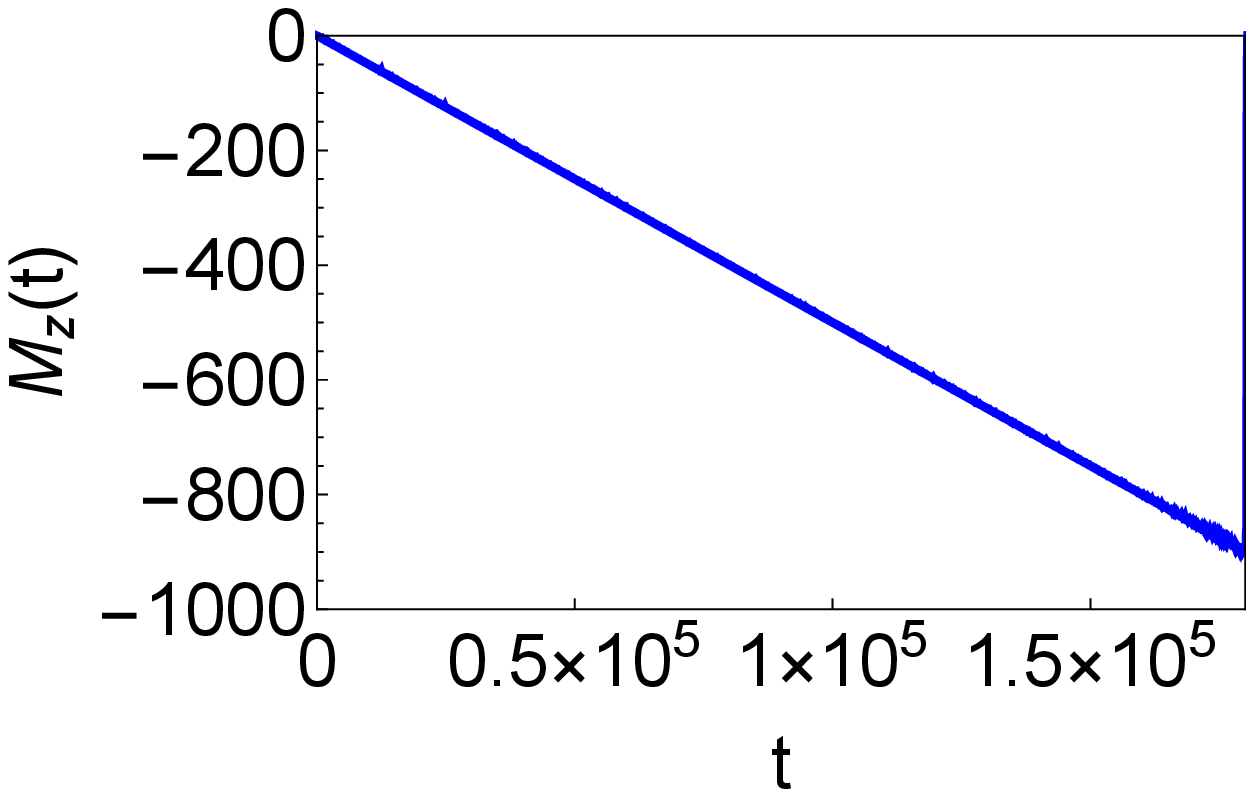,height=3.6cm,width=6.3cm } }
\put(29.5, 20){ \epsfig{file=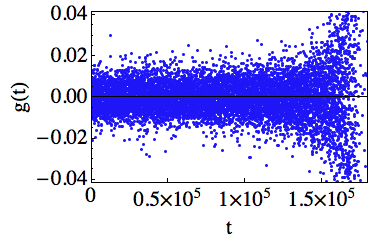,height=2.67cm,width=4.2cm } }
\put(-16, 35){\text{(a)}}
\put(29, 44){\text{(b)}}
}
\end{picture}
\caption{Outcome of the feedback-cooling simulations for $10\times 10\times 10 $ classical spin lattice: (a) the steered variable $M_z = \sum_m S_m^z$ and (b) the feedback field $g(t)$. Fully polarized state corresponds to $M_z = -1000$.}
\label{nmr}
\end{figure}

The results of the simulation are shown in Fig.\ref{nmr}. The value of $|M_z|$ of about 90 percent of the maximum polarization was achieved starting from an unpolarized infinite temperature state, while the feedback field $g(t)$ had a very low amplitude relative to the interaction coefficients in $\mathcal{H}_0$. The divergence of $g(t)$ at the end of the simulated time was an indication that the feedback scheme was about to become unstable.

\section{Theoretical explanation}
\label{explanation}
\begin{figure}[t!] \setlength{\unitlength}{0.1cm}
\begin{picture}(80 , 56 )
{

\put(-7, 0){ \includegraphics[scale=0.47]{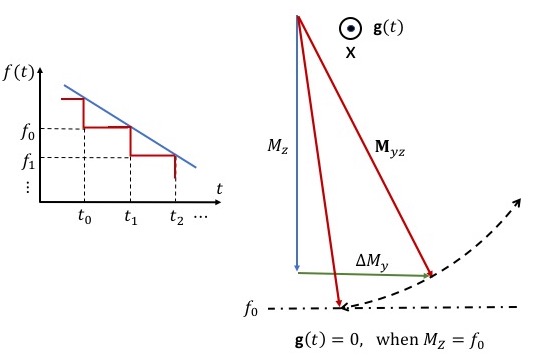}}
\put(-7, 54){(a)}
\put(32, 54){(b)}
}
\end{picture}
\caption{
Conceptual sketch of the feedback control mechanism: (a) Step-wise discretization (red line) of the continuous steering function $f(t)$ (blue line); (b) Evolution  during one time step $[t_0, t_1]$: the $yz$-polarization $\mathbf{M}_{yz}$ is rotated by the driving field $\mathbf{g}(t)$ until $M_z$ becomes equal to $f_0$. }
\label{fig-Mechanism}
\end{figure}

Let us now give the qualitative explanation of the above-reported  cooling effect. The effect capitalizes on the statistical noise of the total spin polarization in the $y$-direction, i.e. in the direction transverse to both the monitored polarization and the feedback field $\mathbf{g}$. Let us for the sake of explanation, discretize the time evolution of the steering function in  steps as shown in Fig.~\ref{fig-Mechanism}(a), such that 
$f(t)$ jumps each time interval $\Delta t$  by $\Delta f = \dot{f} \Delta t$ and then stays constant until the next jump, so that it takes equally spaced values $\{ f_0, f_1, f_2, ... \}$ at respective times $\{ t_0, t_1, t_2, ...\}$. 
Let us further assume that $M_z(t_0) = M_{z0}$, while the value of the steering function has just jumped to $f_0 = M_{z0} + \dot{f} \Delta t$. [Note that in our simulations both $M_z$ and $\dot{f}$ are negative.]
We now observe that the total spin polarization of the system in the $yz$-plane, $\mathbf{M}_{yz}$, has not only the projection $M_z$ on the $z$-axis but also the projection $\Delta M_y $ on the $y$-axis associated with equilibrium fluctuations. Thus $|M_{yz}| = \sqrt{M_z^2 + \Delta M_y^2}$ is slightly larger than $|M_z|$ and also it points at angle $\Delta \phi \approx |\Delta M_y/M_z|$ with respect to the $z$ axis (assuming $|\Delta M_y| \ll |M_z|$). The feedback field $\mathbf{g}(t)$ acts in the $x$-direction, making $\mathbf{M}_{yz}$  rotate back and forth until $M_z(t)$ becomes equal to $f(t) = f_0$ \cite{note-f-Mz}, and then, according to Eq.(\ref{g}), the field switches off. After that, $M_z$ stays equal to $f_0$ until $f(t)$ jumps to $f_1$; then the same mechanism makes $M_z$ reach $f_1$, then $f_2$ and so forth, so that $M_z(t)$ tracks the evolution of the steering function $f(t)$.

In terms of the above discretized description, what are the conditions for our feedback scheme to work? We can identify three of them: 

(i) The jump of the value $\Delta f$ should be smaller than the typical value of $|M_{yz}| - |M_z| \approx \frac{1}{2  } |M_z| \Delta \phi^2$. 

(ii) The time step $\Delta t$ should ideally be larger than the correlation time $T_2$ of the fluctuations of $\Delta M_y$. Otherwise there will be no new statistically independent transverse fluctuation to capitalize on.

(iii) The feedback field $g(t)$ should be large enough  to rotate $\mathbf{M}_{yz}$ by the above-defined angle $\Delta \phi$ during one half of the oscillation period $\pi/\omega$.

Let us apply the above conditions to classical spin lattices. In this case,  $|\Delta M_y| \sim \sqrt{N}$, while we aim at achieving $|M_z| \sim N$. In such a regime, $\Delta \phi \sim 1/\sqrt{N}$, and hence, according to the condition (i), $\Delta f$ is less than a number of order 1. In other words, each time step in the proposed scheme would, at most, increase the {\it total} spin polarization of the system by a number of order 1. This constraint imposes the limitation on the size of the lattices where the {\it relative} polarization $\langle S_z \rangle \equiv |M_z|/N$ of the order of one can be practically achieved. The proposed method would not work for macroscopic systems, because the required number of time steps would be of the order of the Avogadro number. However, the systems consisting of thousands and even millions of spins can be realistically polarized by the method: the maximum number of time steps is then limited by the time $T_1$ characterizing the relaxation of $M_z$ due to the external environment. In the context of NMR, the relaxation time $T_1$ in pure dielectrics can reach $10^3$~s or more, while the transverse relaxation time $T_2$ can  be as small as $10^{-4}$~s.
We also note that, since $M_z = \langle S_z \rangle N$, the polarization increase after each step appearing in condition (i) is proportional to $1/\langle S_z \rangle$. Thus it is noticeably larger for weakly polarized states, which is particularly helpful for NMR. On the other hand, when $\langle S_z \rangle$ approaches 1, the amplitude of the transverse noise decreases. As this happens, any preset value of $\Delta f$ associated with a constant steering rate would become too large, and hence $M_z(t)$ would stop following $f(t)$, rendering the feedback loop unstable.

One might be concerned in the above discussion that $\Delta f$ is smaller that the typical fluctuating local fields with which nuclei act onto each other, and which, therefore, can disrupt the feedback control. Here we note, however, that (i) despite its smallness, the feedback field rotates simultaneously all cluster spins in the same direction, which amplifies its effect, and (ii) the local field fluctuations just cannot change $M_z$, because it is the integral of motion for ${\cal H}_0$.

In terms of our actual simulations with continuous  $f(t)$, the period  of $g(t)$, i.e. $2 \pi/ \omega$,  can be identified as the time step $\Delta t$ for the discretized analog. The feedback field rotates $\mathbf{M}_{yz}$ half the period in one direction and half the period in the opposite one, so that, with the right value of $g_0$, $M_z(t)$ is supposed to reach $f(t)$ during one of the two half periods.   We had $2 \pi/ \omega \approx 1$, while $T_2 \approx 1/3$, hence the condition (ii) requiring $\Delta t \gtrsim T_2$ is satisfied. Condition (i) requires the change of $f(t)$ during time interval $\Delta t$ to be smaller than 1, which, given that $\Delta t \approx 1$ implies that $|\dot{f}| \lesssim 1$. This inequality was conservatively satisfied by the actual value $|\dot{f}| = 0.005$, which helped us to reach the relative polarization of 90 percent. According to condition (iii) with the input from Eq.(\ref{g}),  $\Delta \phi \sim g_0 \Delta M_z \frac{\pi}{\omega}$, where $\Delta M_z \sim \frac{1}{2} M_z \Delta \phi^2$ is the typical value of $|f(t) - M_z|$ implied by condition (i).
Combining the latter two estimates with the assumption $\Delta \phi \sim 1/\sqrt{N}$ and dropping the numerical prefactors, we obtain  the relation $\frac{g_0 \sqrt{N}}{\omega} \sim 1$, which is consistent with our simulation parameters. 

The role of the oscillating factor $\cos \omega t$ in the feedback control function $g(t)$ is to suppress the probability that the feedback field rotates $\mathbf{M}_{yz}$ by large angle in the direction increasing $|f(t) - M_z|$.  Without periodic sign changes of $g(t)$, the feedback field can accidentally drive $M_z$ sufficiently far from $f(t)$, which in turn would lead to the loss of the steering control. On the other hand, for $\omega \gg 1/T_2$, the feedback scheme would not be able to achieve the maximal conversion of the transverse fluctuations $\Delta M_y$ into the the growth of $|M_z|$:  the scheme would either suppress these fluctuations or  lose the steering control over $M_z$.  The former option would then reduce the acceptable values of the steering rate $\dot{f}$.   
Thus the choice of $\omega \sim 1/T_2$ made in the simulations appears to be close to the optimal one.

%{\it Discussion and conclusions.}
\section{ Discussion and conclusions}

Let us now discuss possible generalizations and the improvements of the proposed scheme. One obvious improvement would be to make the feedback parameters $\dot{f}$, $\omega$ and $g_0$ slowly dependent on time such that $|\dot{f}|$ is larger when $\langle S_z \rangle$ is small, and smaller  when $\langle S_z \rangle$  approaches 1. This would accelerate the initial polarization stage, while allowing one to come closer to fully polarizing the system. 

Another more radical modification can involve monitoring $\Delta M_y$ instead of $M_z$ and then applying $g(t)$ in the form of short pulses \cite{quine2010} with the appropriate sign and amplitude, such that $\mathbf{M}_{yz}$ is rotated towards the $z$-axis. One obvious advantage of such an approach is that the required accuracy for measuring $M_z$ is of order 1, while the required accuracy for measuring $\Delta M_y$ is of order $\sqrt{N}$.

Regarding the applicability of the proposed scheme to quantum spin clusters, our preliminary results (not included in this article) indicate that the scheme should work.  We implemented it for a small cluster of spins $1/2$ with monitored variable being the {\it expectation value} of the total magnetization $M_z$. These simulations are yet to be extended to explicitly include the effects of quantum measurements on the system.

The feedback-based cooling technique can be compared with the dynamic nuclear polarisation (DNP) \cite{maly2008}.  The former may be more difficult to implement, but it has the advantage of being applicable to settings where the laser pumping required for DNP is not possible. Other techniques for polarizing and/or purifying small nuclear spin clusters have also been proposed \cite{rugar05,alvarez2010,dasari2021}. The comparison with those techniques should wait until the present proposal is implemented experimentally. It is our hope that it will allow one to achieve high spin polarizations for larger clusters.

Finally, while the proposed feedback scheme was presented for spin systems, it is conceptually applicable to any many-body system having a  collective variable, which is conserved by the internal dynamics but can be changed by an external perturbation.

In conclusion, we introduced and numerically verified a scheme for polarizing and  cooling large but finite clusters of many particles, and we have also presented the initial analysis of the main physical factors that control the efficiency of the proposed scheme. The scheme is specifically tested in a setting representative of NMR experiments.

B.F. acknowledges the support by a grant of the Russian Science Foundation (Project No. 17-12-01587).

\bibliographystyle{apsrev4-1}
\bibliography{CControl}

\begin{thebibliography}{46}
\bibitem [{not({\natexlab{a}})}]{note-energy}%
  \BibitemOpen
  \href@noop {} {} ({\natexlab{a}}),\ \bibinfo {note} {in this case, $M_z$ is
  not correlated with the total energy of the system despite the $S_m^z S_n^z$
  coupling in the Hamiltonian. The reason is that $\sum_n J_{mn}^z = 0$ due to
  the factor $1- 3\cos^2 \theta_{mn}$.}\BibitemShut {Stop}%
\bibitem [{not({\natexlab{b}})}]{note-f-Mz}%
  \BibitemOpen
  \href@noop {} {} ({\natexlab{b}}),\ \bibinfo {note} {strictly speaking,
  $M_z(t)$ would not become equal to $f_0$ but rather approach $f_0$
  exponentially.}\BibitemShut {Stop}%
\end{thebibliography}
\end{document}